\documentstyle[aps,prb,epsfig,multicol,color]{revtex}

\begin{document}
\draft

\title{Possible Pairing Mechanisms of PuCoGa$_5$ Superconductor}

\author{Yunkyu Bang$^{1,2}$, A. V. Balatsky$^{2}$,  F. Wastin$^{3}$, and J. D.
Thompson$^{2}$}

\address{
$^{1}$ Department of Physics, Chonnam National University, Kwangju
500-757, Korea}

\address {$^{2}$Los Alamos National Laboratory, Los Alamos, New Mexico
87545, USA}

\address {$^{3}$European Commission, Joint Research Centre, Institute
for Transuranium Elements, \\ Post office 2340, D-76175 Karlsruhe,
Germany}

\date{\today}
\maketitle

\begin{abstract}
We examine possible pairing mechanisms  of superconductivity in
PuCoGa$_5$ based on spin-fluctuations or phonons as mediating
bosons. We consider experimental data of specific heat C(T) and
resistivity $\rho(T)$ as input to determine a consistent
scattering boson with the superconducting transition temperature
of 18.5K in PuCoGa$_5$. Irrespective to the type of boson, the
characteristic boson frequency is found to be $\sim 150 K$ from
the resistivity fitting. The spin fluctuation model is most
consistent with the experimental resistivity, successfully
explaining  the anomalous temperature dependence ( $\sim
\frac{T^2}{150 K +T}$) at low temperatures as well as the
saturation behavior at high temperatures.
Assuming that the pairing state is non s-wave, the large residual
resistivity $\rho_{imp} \sim 20 \mu \Omega cm \sim 120 K$ suggests
that an ideally pure sample of PuCoGa$_5$ would have a maximum
T$_c$ of 39 K.

\end{abstract}



\begin{multicols}{2}

\section{Introduction}

Recently, superconductivity was found in PuCoGa$_5$ at
the amazingly high transition temperature (T$_c$) of 18.5 K
\cite{Nature}. Considering the fact that the highest T$_c$ of
f-electron based compounds \cite{CeCoIn5} was $\sim$  2K , the
T$_c$ of 18.5 K   is an order of magnitude larger value than the
previous highest T$_c$ in f-electron based superconducting
compounds. Therefore, the understanding the origin of this 18.5
K transition temperature in PuCoGa$_5$ should not only provide
important information on the puzzling behavior of f-electrons but
also shed light on the origin of the high transition temperature
in cuprate superconductors.

Let us briefly review  the experimental data known about
PuCoGa$_5$. First, from the specific heat jump at T$_c$, the Sommerfeld
coefficient is $\gamma_{normal} \sim 77$ mJ/K$^2$mol. Second, from
the T$^3$ phonon contribution in C(T), the typical value of the phonon
frequency scale ($\Theta_D$) is estimated to be 240 K. Third, the
resistivity $\rho(T)$ shows a typical S-shape behavior in its
temperature dependence, which is often observed in spin
fluctuating heavy fermion compounds such as UPt$^3$
[\cite{UPt3}]. Another important piece of information from
$\rho(T)$ is its magnitude; the value $\rho(T=300K)\sim 250 \mu
\Omega$ cm itself indicates strong scattering of the conduction
electrons (an order of magnitude larger than the values of
CeMIn$_5$ (M=Co,Ir,Rh) superconductors having  T$_c$ of $\sim$ 2 K
\cite{CeCoIn5,Hegger}). Forth,  puzzling  is the data of the
uniform susceptibility $\chi(T)$; it shows an almost exact
Curie-Weiss temperature dependence of $\chi(T) \sim
1/(T+T_\theta)$ with $T_\theta=2 K$. This indicates that there are
almost free local moments in the temperature range of 18.5 to 300
K with an effective local moment magnitude (0.68 $\mu_B$) close
to the value of the local moment of free Pu$^{3+}$ (0.84 $\mu_B$)
\cite{ISSP}. We believe that these local moments responsible for
the observed $\chi(T)$ are not coupled (or negligibly weakly
coupled) to the conduction electrons and play no significant role in the
transport properties as well as in the superconducting pairing.
This does not mean that there are no interesting spin fluctuations
-- their contribution to $\chi(T)$ might be much smaller than the
local moment contribution or  more probably their temperature
dependence is not pronounced. If this is the case, the Curie-like
$\chi(T)$ should continue to exist below the transition
temperature and should be observable if the diamagnetic part of
$\chi(T)$ is subtracted or suppressed below T$_c$.

As a possible pairing mechanism and pairing symmetry, we do not
have much decisive data except estimates of a few energy scales.
First of all, conventional phonon mediated pairing seems not
unreasonable but only barely possible with values of
$\Theta_D =240 K$, dimensionless coupling constant $\lambda=0.5
\sim 1.0$, and a typical value for the Coulomb pseudopotential
$\mu^*=0.1$.  The Allen-Dynes' T$_c$ formula \cite{McMillan} T$_c
= \frac{\omega_{ph}}{1.20} \exp [-
\frac{1.04(1+\lambda)}{(\lambda-\mu^*(1+0.62 \lambda))}]$ provides
T$_c$=16.7K for $\lambda=1$ and T$_c$=2.9K for $\lambda=0.5$,
respectively. On the other hand, although there is not yet direct
experimental evidence, the  existence of Pu f-orbitals
participating in Fermi level crossing band(s) \cite{FS} and the
isostructure  CeMIn$_5$ (M=Co,Rh,Ir)  compounds let us
suspect the important role of spin-fluctuations to explain the
normal state transport properties as well as the pairing mechanism
in PuCoGa$_5$.

In this paper, we examine two possible bosonic scattering mechanisms,
namely, phonons and spin-fluctuations, to consistently understand
the available experimental data mentioned above.  Our strategy is
the following. Assuming each bosonic scattering, we try to fit the
dc-resistivity data $\rho(T)$ for its temperature dependence as
well as its magnitude. From this fitting procedure, we extract the
magnitude of the dimensionless coupling constant $\lambda$ and the
typical energy scale of the corresponding boson. From these two
numbers, we then can estimate  T$_c$ using McMillan's formula.

\section{Formalism}

We calculate the conductivity using the Kubo formula as follows.

\begin{equation}
\sigma(T)= \frac{\hbar e^2}{3} \sum_k v^2(k) \int \frac{d
\omega}{4 \pi  T} A^2 (\vec{k},\omega) [\frac{1}{\cosh^2
[\omega/2T]}]
\end{equation}

\noindent
  where $A (\vec{k},\omega)=2 Im G_R (\vec{k},\omega)$ is
the one particle spectral density of quasiparticle in the
conduction band and the retarded Green function of the
quasiparticle is defined as  $G_R (\vec{k},\omega)=
\frac{1}{\omega-\epsilon_p-\Sigma(\vec{k},\omega)}$. All
scattering information is included in the self-energy
$\Sigma(\vec{k},\omega)$. Within the Born approximation,  the
self-energy is calculated as

\begin{equation}
\Sigma(\vec{k},\omega_n)= g^2 T \sum_{\Omega_n,q}  \int d \omega'
\frac{B(q,\omega')}{i \Omega_n - \omega'} \frac{1}{i \omega_n + i
\Omega_n - \epsilon_{k+q}},
\end{equation}

\noindent where $B(q,\omega)$ is the spectral density of a bosonic
propagator $D(q,\omega)=\int \frac{d \omega'}{2 \pi}
\frac{B(q,\omega')}{i \Omega_n - \omega'}$ and $g$ is the
electron-boson coupling constant. After a Matsubara frequency
($\Omega_n$) summation, the imaginary part of $\Sigma_R
(\vec{k},\omega + i \eta)$ is written as

\begin{eqnarray}
Im \Sigma_R (T,\omega + i \eta)&=& g^2 N(0)   \int \frac{d
\omega'}{2 \pi} \\ \nonumber & & \times \sum_{q} \pi B(q,\omega')
[n (\omega') + f(\omega+\omega')]
\end{eqnarray}

\noindent where $n (\omega)$ and $f (\omega)$ are the Boson and
Fermion distribution functions, respectively. $N(0)$ is the
density of states per spin at the Fermi level. Plugging the
self-energy Eq.(2) into Eq.(1) and summing $\sum_{k}$, Eq.(1) gives

\begin{equation}
\sigma(T)= \frac{\hbar e^2}{3} N(0)  <v^2>_{FS} \int \frac{d
\omega}{4 T}  [\frac{1}{\cosh^2 [\omega/2T]}] \frac{1}{Im \Sigma_R
(T, \omega)}
\end{equation}

A few remarks are in order for the above equations. First, the
vertex correction is ignored. The justification is that when the
scattering is local in space -- technically meaning that the
self-energy is momentum independent -- the current  vertex is not
renormalized. This is  consistent with the local approximation in
calculating Eq.(2) and Eq.(3); consistent with this,  we also neglect the
momentum dependence of the coupling $g$ implying every
quasiparticle is equally scattered by the assumed boson.  Second,
the self-energy is calculated only in the Born approximation. Third,
the Fermi surface (FS) anisotropy is neglected, resulting the
factor $\frac{1}{3}$ and the FS averaged Fermi velocity squared
$<v^2>_{FS}$. With all these approximations, we should take the
temperature power law of the calculated resistivity at low
temperatures with reservation. Otherwise it  induces an error of
order O(1) for the overall magnitude. Finally, $N(0)$ and
$<v^2>_{FS}$ are the values before they are renormalized by the
bosonic scattering and, therefore, are difficult to be estimated from
experiments. We rewrite Eq.(4) as follows.

\begin{equation}
\sigma(T)= \frac{\hbar e^2}{3} Z \tilde{N}(0)  <\tilde{v}^2>_{FS}
\int \frac{d \omega}{4 T}  [\frac{1}{\cosh^2 [\omega/2T]}]
\frac{1}{Im \Sigma_R (T, \omega)}
\end{equation}

In the above equation,  $\tilde{N}(0)$ and   $\tilde{v}$ are the
quantities renormalized by the bosonic scattering and $Z$ is the
wave function renormalization parameter $Z=1+\partial Re
\Sigma(\omega) / \partial \omega$.
The above expression is very useful for our purpose. The
renormalized quantities $\tilde{N}(0)$ and   $\tilde{v}$ can be
obtained from experiments. Furthermore, although Eq.(5) has an
explicit dependence on   $Z$, the implicit dependence of $Im
\Sigma(T, \omega)$ on $ Z$ makes Eq.(5) a slowly varying function of
$Z$. The reason is because the real and imaginary parts of $\Sigma
(T,\omega)$ are related by the Kramer-Kronig relation, so that
$\partial Re \Sigma(T, \omega) /
\partial \omega = Z(T)-1$ resulting  $Im \Sigma(T, \omega) \sim
[Z(T)-1]$.

\section{Results}

\subsection{Spin fluctuations}

We choose the mean field type spin relaxational mode of
$B(q,\omega)=\frac{C \omega }{[I(T)+b(\vec{q}-\vec{Q})^2]^2
+[\frac{\omega}{\Gamma}]^2}$, where $\Gamma$ is the magnetic Fermi
energy \cite{Monthoux}, $I(T)=I_0 + a T$ is the parameter
controlling the distance from a magnetic quantum critical point,
and $\vec{Q}$ is a typical wave vector of the magnetic ordering.
$\Gamma \cdot I (T) =\omega_{sf} (T)$ defines the characteristic
energy scale of the fluctuations. $b$ describes the dispersion of
the collective mode of $D(q,\omega)$. The magnetic ordering vector
$\vec{Q}$ can be in two dimensions or in three dimensions
depending on the nature of the incipient magnetic order. This
dimensionality of magnetic ordering would affect the power law of
the resistivity at low temperatures.
We assume 2-D in our calculations, reflecting the band calculations
\cite{FS}. The overall magnitude $C$ of $B(q,\omega)$ is combined
with $g^2 N(0)$ in Eq.(3) to determine the overall magnitude of
$\Sigma(T,\omega)$. This overall magnitude is determined once
$(Z(T)-1)$ is fixed. Therefore, we do not need to determine
separate values of $g^2$, $N(0)$, and $C$. Most importantly, $I_0$
determines the low temperature T$^{\alpha}$  region of the
resistivity before the inflection point; $\alpha \sim 4/3$ was
extracted from experiments \cite{Nature} for T$_c$ $<$ T $<$ 50K,
but a dimensional counting for the spin fluctuations model in 2-D
gives $\rho$(T) $\sim \frac{T^2}{I_0+a T}$ at low temperatures.
The temperature variation of $a T$ in $I(T)$ also controls
the high temperature saturation behavior of $\rho(T)$, since
increasing value of $I(T)$ for larger temperatures reduces the
scattering rate.

Fig.(1) shows the typical results from Eq.(5) with $Z$
values varying from 2 to 8. To be quantitative,  the experimental values of
$\tilde{N} (0) (\leftarrow \gamma=77 mJ/K^2 mol)$  and $\tilde{v}
(= 8.8 \times10^{6} cm/sec)$\cite{ISSP} are used. Given these two
experimental input values, there is no free parameter to adjust
the overall magnitude of the resistivity. The detailed temperature
dependence of $\rho$(T) is controlled by $I(T)$.  $\Gamma \cdot
I_0= \omega_{sf}=$ 150 K, $a=1$ are chosen for illustration. As
seen in Fig.(1), with increasing $Z$, the sensitivity
of $\rho(T)$ to $Z$ becomes weaker, and for all values of $Z$ the
calculated $\rho (T)$ is smaller than the experimental one by a
factor of 3 to 4 (see Fig.2). In order to fit the experimental
$\rho_{exp} (T)$, we tune the value of $\tilde{v}$,
which is a rather rough estimate from the critical magnetic fields
($H_{c2}(0)$) in superconducting state of PuCoGa$_5$, while the
value of $\gamma_{exp}=77 mJ/K^2 mol$ is more reliable.  We also
chose to use $Z=4.6$ which is the ratio of $\gamma_{exp}/
\gamma_{band}$ \cite{FS}. The result is shown in Fig.(2) in
comparison with the experimental $\rho_{exp} (T)$. Input
parameters are $\omega_{sf}=$ 150K, $a$=1, $\gamma=77 mJ/K^2 mol$,
and $\tilde{v}_{exp}=4.78 \times10^{6} cm/sec$. The residual
resistivity $\rho_{imp}=15 \mu \Omega$ cm is added to the
theoretical result (The better fitting of the low temperature part
only would give $\rho_{imp}=19 \mu \Omega$. See the inset of
Fig.(2).) The overall fitting is satisfactory from low to high
temperatures.
In particular, the saturation behavior at high temperatures is
well reproduced with the temperature dependent I(T). This is an
expected behavior since increasing temperature should shorten the
magnetic correlation length such as $\xi^{-2} (T) \sim (I_0 + a
T)$. Some discrepancy between the theoretical result and the
experimental one beyond 250 K (Fig.2) would be due to the failure
of the simple relation $\xi^{-2} (T) \sim (I_0 + a T)$ at such
high temperatures.
To obtain better fitting in the low temperature region, one needs
to allow a modification of the high temperature part. The result
is shown in the inset. It indeed reproduces the experimental
observation $\rho$(T) $\sim$ T$^{4/3}$ [\cite{Nature}] for T$_c <$
T $<$ 50 K, while the correct theoretical form of  $\rho$(T) at
low temperatures  is $\sim \frac{T^2}{150 K +T}$.

\begin{figure}
\epsfig{figure=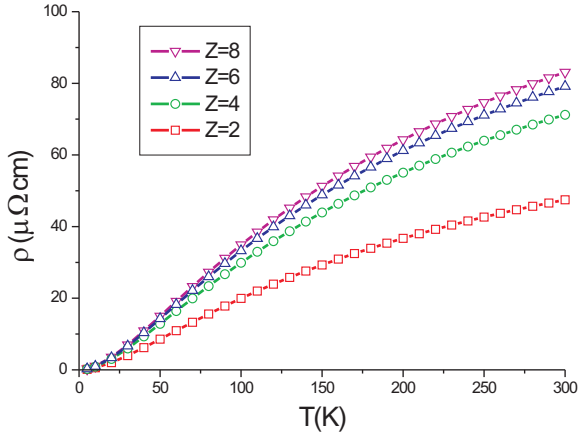,width=1.0\linewidth} \caption{Theoretical
calculations of  resistivity $\rho(T)$ for varying $Z$ values ($Z$
=2,4,6,8). Input parameters are $\omega_{sf}=$ 150K, $a=1$ ,
$\gamma=77 mJ/K^2 mol$, and $\tilde{v}_{exp}=8.8 \times10^{6}
cm/sec$. \label{fig1}}
\end{figure}

\begin{figure}
\epsfig{figure=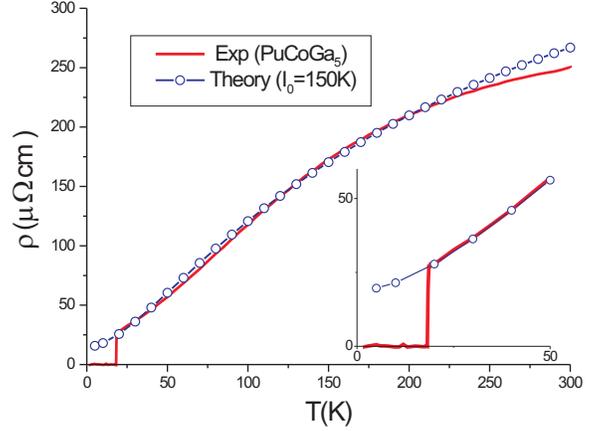,width=1.0\linewidth}
\caption{Theoretical calculation with $Z=4.6$ (open blue circles)
and the experimental resistivity $\rho(T)$\cite{Nature} (red solid
line). Input parameters are $\omega_{sf}=$ 150K, $a=1$, $\gamma=77
mJ/K^2 mol$, and $\tilde{v}_{exp}=4.78 \times10^{6} cm/sec$ and
$\rho_{imp}=15 \mu \Omega$cm is added. Inset: closeup view of low
temperature region. For better fitting, $\tilde{v}_{exp}=5.28
\times10^{6} cm/sec$  and $\rho_{imp}=19 \mu \Omega$cm are used
\label{fig2}}
\end{figure}

\begin{figure}
\epsfig{figure=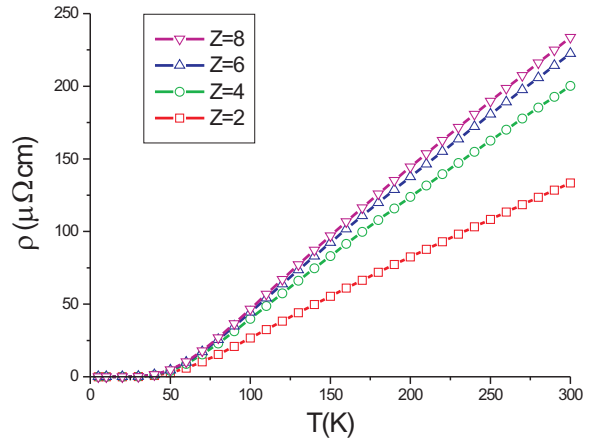,width=1.0\linewidth} \caption{Theoretical
calculations of  resistivity $\rho(T)$ with Einstein phonon with
$\theta_D =240$K for varying $Z$ values ($Z$ =2,4,6,8). Input
parameters are $\gamma=77 mJ/K^2 mol$, and $\tilde{v}_{exp}=8.8
\times10^{6} cm/sec$. \label{fig3}}
\end{figure}

\begin{figure}
\epsfig{figure=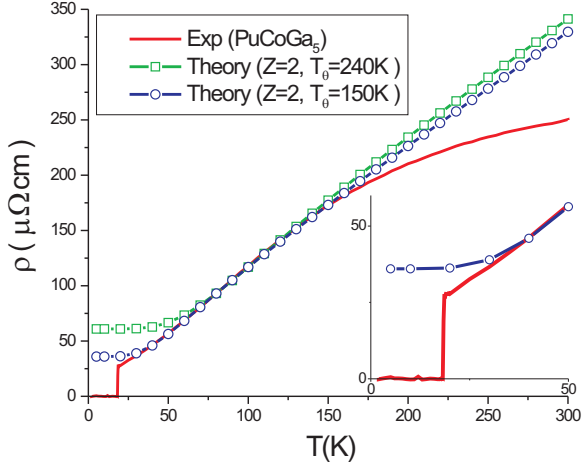,width=1.0\linewidth}
\caption{Theoretical calculations of  resistivity $\rho(T)$  for
$Z=2$  with two different  Einstein phonon frequencies: $\theta_D
=240$K, $\gamma=77 mJ/K^2 mol$, and $\tilde{v}_{exp}=6.1
\times10^{6} cm/sec$ (open green squares);  $\theta_D =150$K, and
$\tilde{v}_{exp}=5.6 \times10^{6} cm/sec$ (open blue circles). The
experimental resistivity $\rho(T)$ \cite{Nature} is red sold line.
Inset: closeup view of low temperature region.  \label{fig4}}
\end{figure}

\subsection{ Phonons}

Fig.3 shows the resistivity calculated with an Einstein phonon
$B(q,\omega) \sim \delta (\omega-\theta_D) $ with $\theta_D=240
K$, which is the value obtained from the specific heat C(T) above
T$_c$ \cite{Nature}, for varying values of Z from 2 to 8. The
input values are $\gamma=77 mJ/K^2 mol$  and $\tilde{v} (= 8.8
\times 10^6 cm/sec)$.
As in the spin-fluctuations case (Fig.1), with increasing Z values the
sensitivity  of $\rho(T)$ to Z becomes weaker, and the overall
magnitude of $\rho_{theor}(T)$ is smaller than $\rho_{exp}(T)$ by
the factor of 2 to 3 (compared the values at T=100K). In Fig.4, we
show the tuned theoretical results of $\rho_{theor}(T)$ in
comparison with the  experimental $\rho_{exp} (T)$. As before, in
order to increase the overall magnitude of $\rho_{theor}(T)$, we
tune the renormalized Fermi velocity $\tilde{v}$.

In Fig.4, we see that the theoretical resistivity
$\rho_{theor}(T)$ with $\theta_D =240 K$ (green open squares) has
a higher power law region over a braoder range of low temperatures relative
to the experimental data. For overall fitting of the
data, we add a large impurity resistivity $\rho_{imp}=61 \mu
\Omega$ cm, which makes the low temperature part in clear
disagreement with experiment.  To fit the low temperature region
better, we need to reduce $\theta_D$. The blue open circles are
the result with $\theta_D=150 K$. This happens to be the same
bosonic energy scale as used in the spin-fluctuation model
fitting. It means that the resistivity data reveal a
characteristic energy scale of the scattering boson to be $\sim$ 150 K,
irrespective of the origin of the boson. An impurity resistivity
$\rho_{imp}=36 \mu \Omega$ cm is added. The inset shows a
close-up view of the low temperature region. There is a clear
deviation between $\rho_{ph}$(T) $\sim \exp{[-\theta_D /T]}$ and
$\rho_{exp}$(T) $\sim$ T$^{4/3}$ [ \cite{phonon}].

The phonon model has a critical defect at high temperatures with
any reasonable parameters. While the experimental $\rho_{exp}(T)$ show a
clear saturation behavior for T $>$ 160 K, phonon scattering
results in an ever increasing T-linear resistivity for
temperatures beyond a fraction of $\theta_D$. This saturation
behavior in resistivity is a long standing problem in
A-15 compounds \cite{resistivity_sat}, transition metals, and some
heavy fermion compounds such as UPt$^3$ [\cite{UPt3}]. While there
is no general mechanism to explain this phenomena
\cite{Allen_sat}, we can make the following remark.

With the spin-fluctuation model, this saturation behavior is
naturally explained by the temperature dependence of the magnetic
correlation length $\xi ^{-2}(T) \sim I(T)$ (see Fig.2).
On the other hand, phonon scattering  needs to invoke a
separate mechanism to explain the saturation behavior. Recently,
Calandra and Gunnarsson \cite{Gunnarsson}, assuming lattice
vibrations couple with the electron hopping integral (HI), have shown
that the saturation resistivity arises at high temperatures due to
a cancellation between an increasing phonon population and an
increasing electron kinetic energy. Without a specific model and
numerical calculations, this saturation behavior can be described
phenomenologically in a two parallel resistor model ("shunting
model") as \cite{Wiesmann}

\begin{equation}
\rho^{-1}(T)= \rho^{-1} _{ideal}(T) + \rho^{-1} _{max},
\end{equation}

\noindent where $\rho _{ideal}(T)$ = $\rho _{e-ph}(T)+
\rho_{imp}$, and $\rho _{e-ph}(T)$ is calculated by Eq.(5). $\rho
_{max}$ is the maximum resistivity determined by the $f$-sum rule
\cite{Gunnarsson}, but here determined by empirical fitting.

\begin{figure}
\epsfig{figure=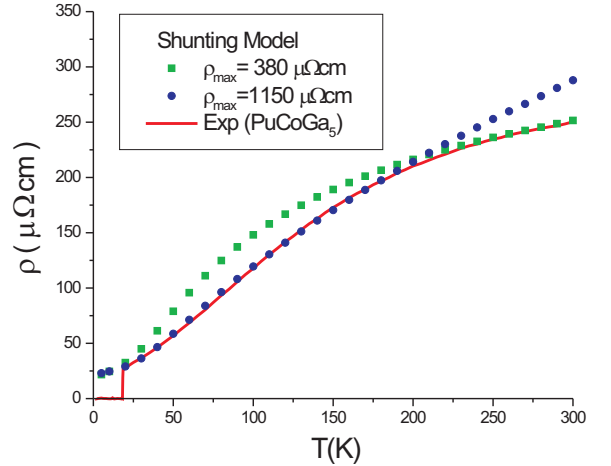,width=1.0\linewidth} \caption{Shunting
model fits to the experimental resistivity (red sold line). The
low temperature fit (open blue circles) is with
$\tilde{v}_{exp}=4.9 \times10^{6} cm/sec$ and $\rho_{max}=850 \mu
\Omega$ cm. The high temperature fit (open green squares) is with
$\tilde{v}_{exp}=3.6 \times10^{6} cm/sec$ and $\rho_{max}=380 \mu
\Omega$ cm. \label{fig5}}
\end{figure}

In Fig.5, we show the best fit results with this model. For the
electron-phonon resistivity $\rho_{ph}(T)$, we use Z=2 and
$\theta_D =150 K$ as in Fig.4 and the overall magnitude is tuned
with $\tilde{v} _{exp}$. As seen, the simple shunting model cannot
fit the whole temperature region with any parameters in contrast
to the successful cases of A-15 compounds \cite{Wiesmann}. The
implication is  either that the shunting model is not good enough
to fit the fine details or that phonon scattering is simply
not the correct model for the resistivity of PuCoGa$_5$.

\subsection{ T$_c$}

Before we estimate T$_c$, we need to consider the effect of
impurities in the sample. The effect of impurities on $T_c$
depends on the gap symmetry. For an s-wave gap -- in the case of
phonon  pairing  --  T$_c$ is not reduced by nonmagnetic
impurities \cite{Anderson} in the first approximation. However,
for an unconventional non s-wave gap -- in the case of spin
fluctuation pairing  -- T$_c$ is reduced by any type of
impurities.
As noticed in the previous section, the experimental resistivity
has a large residual value of $\rho_{imp} \sim 20 \mu \Omega
cm$, which should be a serious pair breaker for non s-wave
pairing.

Assuming $\rho_{imp} = 20 \mu \Omega cm$ in the spin fluctuation
model and  plugging Z=4.6 and $\tilde{v} _{exp} = 4.78 \times
10^6 cm/sec $ in Eq.(5), we obtain $Im \Sigma_{imp} =\Gamma_{imp} =
116 K $. This scattering rate is far larger than T$_c$=18.5 K.
From the theory of Abrikosov-Gor'kov \cite{AG}, the transition
temperature is reduced by $\sim \frac{\pi \Gamma_{imp}}{4}$.  But
the original Abrikosov-Gor'kov formula should be modified by the
mass renormalization factor Z as  $\frac{\pi \Gamma_{imp}/Z}{4}$.
With Z=4.6 and $\Delta$ T$_c$ =T$_{c0} -$ T$_c \sim  20 $ K, we
estimate the transition temperature of an ideally pure sample to
be T$_{c 0} \simeq$ 39 K.
Therefore, it appears that if the gap symmetry of PuCoGa$_5$
is unconventional, as in most of heavy
fermion superconductors, PuCoGa$_5$ is another true high
temperature superconductor. The important question is then whether
the parameters of the spin fluctuation model obtained in the previous
section can produce a transition temperature of $ \sim 39$  K. For
a ballpark estimate, we use the Allen-Dynes
formula\cite{McMillan} for T$_c$,

\begin{equation}
T_c= \frac{<\omega>}{1.20} \exp[-
\frac{1.04(1+\lambda)}{\lambda}],
\end{equation}

\noindent assuming $\mu^{*} =0$ for non s-wave pairing channel.

The important parameter is $<\omega>=[\int_0 ^{\infty} d \omega~
\alpha^2 B(\omega)]/[\int_0 ^{\infty} d \omega~ \alpha^2
B(\omega)/ \omega]$, which  defines the characteristic energy
scale of the pairing boson. The nominal characteristic energy
scale of spin fluctuations is $\omega_{sf}=150$ K, but the above
Allen-Dynes definition of $<\omega>$ would give several times $\omega_{sf}$
because of the $1/ \omega$ long tail of the spin
fluctuation spectra $B(\omega)$. In reality, the effective cut
off energy scale should be between these values, say,  $3 \sim
4 ~\omega_{sf}$ \cite{Monthoux}. Without solving the strong
coupling T$_c$ equation with details of the band structure and
full dynamics of the spin fluctuations, we cannot tell the precise
value of this.
With this reservation for $<\omega>$, the Allen-Dynes formula with
$\lambda = Z - 1 = 3.6$ and $<\omega> \sim \omega_{sf} =150$ K
indeed produces T$_c \sim 35$ K.

For the phonon case, nonmagnetic potential scatterers do not
affect T$_c$ for an s-wave symmetry gap\cite{Anderson}. Therefore,
we do not need to consider impurity effects on T$_c$.  The
Allen-Dynes formula with $\lambda = 1$, and $\mu=0.1$ gives T$_c$=
10.44 and 16.7 K for $<\omega>$=150 K and 240 K, respectively.
However, if we assume that  the total mass renormalization (Z=4.6)
is caused solely by phonon scattering, the effective coupling
$\lambda \sim 3.6$ should be used, and it would be strong enough to
produce T$_c > 20 $ K  with $<\omega>$=150 K.

\subsection{Natural radiation damage}
PuCoGa$_5$ should have radiation-induced self damage as a function of   time.
A preliminary measurement \cite{Nature} indicates a decrease
of T$_c$ at a rate of $\sim$ 0.2 K/month, due to a radiation
damage, which appears to be a quite slow suppression rate at a first
look. Here, we estimate a theoretical T$_c$
suppression with radiation damage and compare it with experiments.
Because there is no study of radiation damage in PuCoGa$_5$
itself, we use results from Pu metal ($\delta$-phase)\cite{Wolfer}.
%
%
%
%
%
Each Pu decays into U and an $\alpha$ particle. Both particles
collide with nearby Pu nuclei creating so-called Frenkel pairs
consisting of a vacancy and a self-interstitial of Pu ions. In the
case of $\delta$-Pu, the total number of Frenkel displacements
from one Pu is 0.1033 per year. Given these data, the estimated
number of displacements  per Pu per month is $\sim$ 0.86 $\%$.
Assuming the unitary limit of the impurity scattering strength,
the impurity scattering rate is given by $\Gamma_{imp} =
\frac{n_{imp}}{\pi N(0)}$, and therefore $\Gamma_{imp} \sim 2 $ K.
For s-wave pairing, this little potential scattering should have
no effect on T$_c$ suppression. However, it is not certain whether
the displaced Pu ions would behave as  potential scatterers or
magnetic scatterers. For non s-wave unconventional pairing,
T$_c$ decreases even with potential scatterers. With a modified
Abrikosov-Gor'kov formula, $\Delta$ T$_c$ =T$_{c0} -$ T$_c$=
$\frac{\pi \Gamma_{imp}/Z}{4}$, we obtain $\Delta$ T$_c \sim 0.3$
K per month, assuming the mass renormalization factor $Z=5$,
consistent with the experiment. Therefore, we conclude that the
T$_c$ degradation with natural radiation damage is consistent both
with unconventional pairing  and also with s-wave pairing, if
the displaced Pu ions behave as magnetic impurities in the latter
case.

\section{Conclusion:}

We have found rather specific constraints on possible pairing bosons to
be consistent with available experimental data for PuCoGa$_5$. We
examined two possible bosons as a common source for the normal
state resistivity and  the superconducting pairing.

For a spin fluctuation model, we found that the characteristic spin
fluctuation energy of 150 K is consistent with the resistivity
data. Theoretical calculations of resistivity produce a
satisfactory agreement with the experimental resistivity: both for
the anomalous power law ($\sim$ T$^{4/3}$) at low temperatures and
for the saturation behavior at high temperatures. The same spin
fluctuations would produce unconventional superconducting pairing, such as
d-wave symmetry. In this case, the large residual resistivity
$\rho_{imp} \simeq 20 \mu \Omega cm$ acts as a serious pair
breaker. To survive with T$_c$=18.5 K after this pair breaking
effect, the original T$_{c0}$ should be $\sim 39$ K. The
Allen-Dynes T$_c$ formula can support this high T$_{c0}$ with
$<\omega> \sim \omega_{sf}$. If this is the case, we have observed
another unconventional high temperature superconductor after the
cuprate superconductors. Perhaps this large impurity effect
explains why UCoGa$_5$ ($\rho_{imp} \simeq 20 \mu \Omega$ cm) is
not a superconductor \cite{U115}. Taken together, we think that
spin fluctuations are the most likely source of superconducting pairing as
well as the normal state resistivity in PuCoGa$_5$.

For the phonon model, we found that the most consistent phonon
frequency to fit the resistivity data should be also 150 K, having
a discrepancy with the estimate  from the specific heat
measurement ($\theta_{D}=240$ K). However, theoretical
calculations of the resistivity with phonon scattering have crucial
defects to explain the experimental resistivity. First, it is
impossible to produce $\rho_{exp}$(T) $\sim$ T$^{4/3}$ with phonon
scattering at low temperatures. Second, the saturation behavior at
high temperatures needs a separate explanation -- this high
temperature feature in resistivity has no direct relation with the
superconducting pairing mechanism, though. Putting aside these
defects and pushing the possibility of the phonon pairing, the
merit of phonon mediated s-wave pairing is its insensitivity to
large impurity scattering observed in the sample. Using a
typical strong coupling constant value $\lambda=1$, it is
difficult to achieve T$_c$ of 20 K. But, assuming that phonon
scattering is the main source  of the large mass renormalization
of Z $ \sim 4.6$, the Allen-Dynes T$_c$ formula with
$<\omega>=150$ K can  easily support T$_c$ of $\sim$ 20 K. In this
case we find some similarity with the phonon mediated A-15
compound superconductors, such as Nb$_3$Sn.
However, comparing with the spin fluctuation model, we think that
the phonon model is unlikely in PuCoGa$_5$. Some key questions to
be answered for the phonon model are: (1) the experimental
$\rho(T) \sim T^{4/3}$ for T$_c$ $<$ T $<$ 50 K [\cite{Nature}] is
not easily reconciled with a phonon scattering mechanism; (2) the
absence of superconductivity in UCoGa$_5$ also needs an
explanation if phonons are the pairing boson.

\section{Acknowledgements}

We thank  M.J. Graf, J. Sarrao, and N. Curro for discussions. Work at Los
Alamos was performed under the auspices of the US DOE. Y.B. was partially
supported by the
Korean Science and Engineering Foundation (KOSEF) through the
Center for Strongly Correlated Materials Research (CSCMR) (2003)
and through the Grant No. 1999-2-114-005-5.

\end{multicols}


\begin{references}


\bibitem{Nature}
J. L. Sarrao, L. A. Morales, J. D. Thompson, B. L. Scott, G. R.
Stewart, F. Wastin, J. Rebizant, P. Boulet, E. Colineau, and  G.
H. Lander, Nature   {\bf 420}, 297 (2002); J. L. Sarrao,  J. D.
Thompson, N. O. Moreno, L. A. Morales, F. Wastin, J. Rebizant, P.
Boulet, E. Colineau, and  G. H. Lander, J. Phys. Condens. Matters
{\bf 15}, s2279 (2002).

\bibitem{CeCoIn5}
C. Petrovic, P.G. Pagliuso, M.F. Hundley, R. Movshovich, J.L.
Sarrao, J.D. Thompson, Z. Fisk, p. Monthoux, J. Phys.: Condens.
Matter {\bf 13}, L337 (2001); R. Movshovich, M. Jaime, J. D.
Thompson, C. Petrovic, Z. Fisk, P. G. Pagliuso, and J. L. Sarrao,
Phys. Rev. Lett. {\bf 86}, 5152 (2001).

\bibitem{UPt3}
G.R. Stewart, Z. Fisk, J.O. Willis, and J.L. Smith, Phys. Rev.
Lett. 52, 679 (1984).

\bibitem{Hegger}
H. Hegger, C. Petrovic, E. G. Moshopoulou, M. F. Hundley, J. L.
Sarrao, Z. Fisk, and J. D. Thompson , Phys. Rev. Lett. {\bf 84},
4986 (2000).

\bibitem{ISSP}
J.D. Thompson, J.L. Sarrao, L.A. Morales, F. Wastin, and P.
Boulet, unpublished (2003).



\bibitem{McMillan}
W.L. McMillan, Phys. Rev. {\bf 167}, 331, {1968}; P.B. Allen and
R.C Dynes, Phys. Rev. B {\bf 12}, 905, {1975}.

\bibitem{FS}
I. Opahle and P. M. Oppeneer, Phys. Rev. Lett. 90, 157001 (2003);
T. Maehira, T. Hotta, K. Ueda, and A. Hasegawa, Phys. Rev. Lett.
90, 207007 (2003).

\bibitem{Monthoux}
P. Monthoux, A.V. Balatsky, and D. Pines, Phys. Rev. Lett. 67,
3448 (1991).

\bibitem{phonon}
Instead of using Einstein phonon spectra, using a Debye phonon
spectra it would produce $\rho_{ph} (T) \sim$ T$^5$ law at low
temperatures \cite{Ashcroft}, again deviating from the experiment.
The high temperature behavior is the same with either phonon
spectra.

\bibitem{Ashcroft}
N. W. Ashcroft and N. D. Mermin, {\it Solid State Physics},
Saunders College Publishing, p 526 (1976).


\bibitem{resistivity_sat}
Z. Fisk and G.W. Webb, Phys. Rev. Lett. 36, 1084 (1976);  P.B.
Allen, W.E. Pickett, K.M. Ho, and M.L. Cohen, Phys. Rev. Lett. 40,
1532 (1978).




\bibitem{Allen_sat}
P.B. Allen, Physica B, {\bf 318}, 24, (2002).


\bibitem{Gunnarsson}
M. Calandra and O. Gunnarsson, Phys. Rev. B {\bf 66}, 205105
(2002).

\bibitem{Wiesmann}
H. Wiesmann, M. Gurvitch, H. Lutz, A. Ghosh, B. Schwarz, M.
Strongin, P. B. Allen, and J. W. Halley, Phys. Rev. Lett. 38, 782
(1977).

\bibitem{Anderson}
P. W. Anderson, Phys. Rev. Lett. 3, 325 (1959).

\bibitem{AG}
A.A. Abrikosov and L.P. Gor'kov, Sov. Phys. JETP, 12, 1243 (1961)
(J. Exptl. Theoret. Phys. (USSR) 39, 1781 (1960)).


\bibitem{Wolfer}
W. G. Wolfer, Los Alamos Science No. 26, p 274 (2000).

\bibitem{U115}
UCoGa$_5$ \cite{Nature} has $\rho$(T) $\sim T^2$ at low
temperatures and $\gamma \sim 10$ mJ/mol K$^2$. Within the spin
fluctuations model, consistently UCoGa$_5$ has a much weaker
coupling and a larger I$_0 >$ 250K.



\end{references}
\end{document}